\documentclass[aps,prb,superscriptaddress,twocolumn,showpacs]{revtex4}
\usepackage{graphicx}

\newcommand{\be}{\begin{equation}}
\newcommand{\ee}{\end{equation}}
\newcommand{\bea}{\begin{eqnarray}}
\newcommand{\eea}{\end{eqnarray}}

\begin{document}

\title{Dipolar ground state of planar spins on triangular lattices}

\author{Paolo Politi}
\email{Paolo.Politi@isc.cnr.it}
\affiliation{Istituto dei Sistemi Complessi,
Consiglio Nazionale delle Ricerche, Via Madonna del Piano 10,
50019 Sesto Fiorentino, Italy}
\affiliation{School of Physics, University of Western Australia,
35 Stirling Highway, Crawley WA 6009, Australia}

\author{Maria Gloria Pini}
\affiliation{Istituto dei Sistemi Complessi,
Consiglio Nazionale delle Ricerche, Via Madonna del Piano 10,
50019 Sesto Fiorentino, Italy}

\author{R. L. Stamps}
\affiliation{Istituto dei Sistemi Complessi,
Consiglio Nazionale delle Ricerche, Via Madonna del Piano 10,
50019 Sesto Fiorentino, Italy}
\affiliation{School of Physics, University of Western Australia,
35 Stirling Highway, Crawley WA 6009, Australia}

\begin{abstract}
An infinite triangular lattice of classical dipolar spins is usually considered to have 
a ferromagnetic ground state. We examine the validity
of this statement for finite lattices and in the limit of large lattices. We find that
the ground state of rectangular arrays is strongly dependent on size and aspect ratio.
Three results emerge that are significant for understanding the ground state properties:
i)~formation of domain walls is energetically favored for aspect ratios below a critical value;
ii)~a vortex state is energetically favored in the thermodynamic limit of 
an infinite number of spins, but nevertheless such a configuration 
may not be observed even in very large lattices if the aspect ratio is large;
iii)~finite range ($R$) approximations to actual 
dipole sums may give spurious results and the limit $R\to\infty$
depends on the way it is taken. For the usual, isotropic limit,
the ferromagnetic state is linearly unstable 
and the domain wall energy is negative for any finite range cutoff.
\end{abstract}


% 75.10.Hk Classical spin models
% 75.75.+a  Magnetic properties of nanostructures
% 75.60.Ch  Domain walls and domain structure
% 77.80.-e  Ferroelectricity and antiferroelectricity

\maketitle

Electric or magnetic fields generated by point dipole sources have significant effects
on polarizable materials. The effects can dominate in thin film and nanostructured
elements. In one and two dimensions, long range order of electric or 
magnetic polarizations in dipole lattices is possible\cite{Malozovsky_1991} 
and it is particularly interesting in view of the very strong impact
that lattice geometry can have on the nature of the long range order.\cite{Rozenbaum_1984} 
Interest in these effects has been renewed by the ability to create 
arrays out of polarizable materials\cite{Hehn_Science_1996,Pescia_Science_1998, Cowburn_NJP_1999} 
with element sizes that can behave to reasonable approximation as point dipole sources.

Whereas long range ordering in infinite two dimensional lattices has been examined in some
detail for different rhombic lattices,\cite{Rozenbaum_1991,Fraerman_1999} 
there is to date very little known about how exactly long range order 
is affected by lattice size, shape and dimension in the transition region between
small and infinite lattices. We find that this is a non-trivial point for two dimensional lattices,
and note that many real systems of technological interest fall in the very large, yet non-infinite, 
category. Our results impact not only the nature of ordering in large, rectangular lattices, but
also some inadequacies and failures in 
the use of an interaction cutoff for the dipolar interaction
(see also Refs.~\onlinecite{Vedmedenko_2000,Fazekas_2003}). 

In this manuscript we study a two-dimensional triangular lattice
of classical planar spins with long range dipolar coupling between spins.
The nearest neighbour distance between spins is taken 
as unit length.
The total dipolar energy is
\be
E = \Omega\sum_{(ij)} {1\over r_{ij}^3} \left[
\vec S_i\cdot\vec S_j -3 { (\vec S_i\cdot\vec r_{ij})
(\vec S_j\cdot\vec r_{ij})\over r_{ij}^2 } \right ] ,
\label{eq_energy}
\ee
where $(ij)$ is a couple of spins and $r_{ij}$ is their distance. The energy
can also be written in terms of an effective field $\vec H_i$
acting at a lattice site $i$ as
$E=-{\Omega\over 2} \sum_i \vec S_i \cdot \vec H_i$.
Each $\vec S_i$ may represent either a single spin or the effective spin
of a single domain magnetic particle. It has been shown\cite{Politi_2002} 
that near-field corrections due to the finite size of magnetic particles can be accounted for
with correction terms that do not alter the general qualitative features of the planar
array interactions.
The prefactor $\Omega={1\over 2} g^2\mu_B^2  S^2$  is positive and absorbs
the modulus of the spin (so that $|\vec S_i|=1$). The prefactor
can be set to $\Omega=1$, if there are no other terms in the energy.

The dipolar interaction between two spins $\vec S_1$ and $\vec S_2$
located at sites $\vec r_1$ and $\vec r_2$ is minimized by aligning them
ferromagnetically along the line $\vec r_{12}=\vec r_1 -\vec r_2$. 
A single chain of spins will therefore order ferromagnetically 
along the chain.  For parallel chains of spins, the field
$\vec H(d)$ generated by an infinite chain at a point some distance $d$ away from
the chain axis plays a special role because $\vec H(d)\equiv 0$
in the continuum approximation.\cite{Politi_2002}  For a discrete
lattice, $\vec H(d)$ does not vanish, but it decays exponentially\cite{BM,DeBell_2000}
with the distance, with a sign depending on the precise discrete structure. 

As a consequence of that,\cite{Rozenbaum_1984,Malozovsky_1991} 
geometry becomes especially important in two dimensions: the sign of $\vec H(d)$
depends on the lattice structure and is negative for a square lattice
but positive for a triangular lattice.\cite{Politi_2002} Spin chains are ferromagnetically
coupled in a triangular lattice and antiferromagnetically coupled in a
square lattice.

This picture breaks down when the chains are finite in length.
Consider a rectangular sample of sides $L_x$ and $L_y$ with aspect ratio
$r=L_x/L_y\ge 1$. We shall show, using direct 
numerical evaluation of Eq. (\ref{eq_energy}), that
the energy of a domain wall parallel to the $x$ axis scales with $L_x$ and becomes negative for
$r<r_c$ where the critical aspect ratio is $r_{c}\approx 4.8$. This implies that domain wall formation
is favored in rectangles with relatively small aspect ratios. Most significantly,
increasing the size of the sample while keeping $r$ fixed, does not
increase the number of domain walls. Whereas one might anticipate that
array shape can drive domain formation as it does in exchange coupled ferromagnets,
the number of domains in a dipole ferromagnet is size independent. This feature is not
found in exchange coupled ferromagnets. The difference is
because the non-local dipole interaction in a dipole array produces both the magnetic anisotropy and the
magnetic coupling required for formation of a domain boundary wall. In the exchange coupled ferromagnet,
exchange interactions are usually negligible beyond nearest or next-nearest neighbors.

In the following we shall discuss these shape and size effects, and also show that vortex formation is energetically
favored only for very large lattices when $r$ is large. 
Appearance of vortices is a well known fact, both in purely dipolar
systems\cite{Belobrov_1985,Vedmedenko_1999} and in dipole plus exchange
systems,\cite{Allenspach_2004} but we are
able to evaluate the vortex energy as a function of size and shape of the
sample.  
Effects of local quadratic and quartic magnetic anisotropies
are also discussed.

We begin by reviewing two results which are particularly enlightening as to why
ferromagnetic lines and vortices are useful concepts for understanding ordering
in two-dimensional dipolar lattices.
Let us consider a closed curve $\Gamma$. If the spin is 
treated as a continuous field, we have $\vec S(\vec r)$ at each point $\vec r$ of
the curve. The first result is that,
if $\vec S$ is tangent to $\Gamma$, the dipolar field
$\vec H(P)$ generated in any point $P$ not belonging to the curve vanishes.\cite{note_first-result}
This result shows that concentric
circles in a vortex or circles belonging
to neighbouring vortices
do not interact for continuous dipole sources.
If the discrete lattice structure is considered, this is no
more exact,
but the interactions are weak, specially for non concentric circles,
as it occurs in neighbouring vortices.

The second result is that
the field produced by all spins of a closed curve $\Gamma$
in a point $P$ belonging to the curve is tangent to
the curve itself.\cite{note_second-result}
Therefore the vortex is the spin configuration
that minimizes the dipolar energy of a curve of spins, as
the ferromagnetic configuration minimizes the dipolar energy of a straight line of
spins.

Two conclusions follow. First, ferromagnetic lines or  vortices 
are energetically favorable because in such configurations
each spin is aligned along the dipolar field. Second, interaction
between (infinite) straight lines or vortex lines is weak (in a
continuum approximation it is exactly zero) and exists only in discrete lattices.
The type of alignment therefore depends not only on size and shape of an array,
but also on the local symmetry defining the geometry of the array.

\begin{figure}
\includegraphics[width=8cm,angle=0,clip=true]{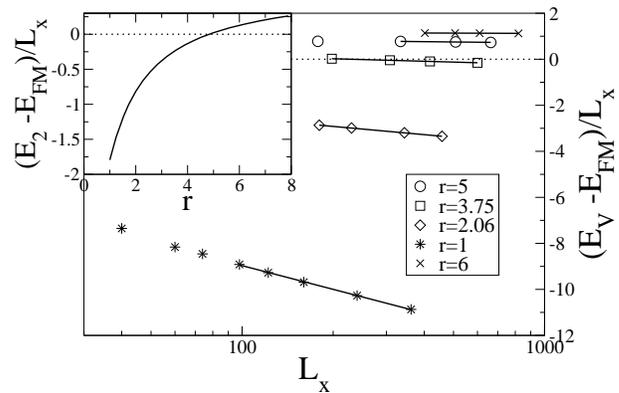}
\caption{Inset: 
difference between the energy of the two-domain state, $E_2$, 
and the energy of the single-domain state, $E_{FM}$,  per unit length
$L_x$ of a rectangular sample of sides $L_x,~L_y$,
as a function of the aspect ratio $r=L_x/L_y$.
The plotted quantity is also called domain wall energy, $E_{dw}$.
Main: energy of a vortex ($E_V$) as a function of the horizontal size $L_x$.
Again, we scale with respect to $E_{FM}$ and per unit length.
Straight lines are fits to the equation $(E_V-E_{FM})/L_x=
a(r)-b(r)\ln L_x$.}
\label{fig_en}
\end{figure}

Now let us
consider a rectangular sample of sides $L_x, L_y$ with $L_x\ge L_y$. 
Because of shape effects,\cite{Politi98}
the ferromagnetic state is directed along the $x$ axis.
We create a domain wall along the $x$ axis and form two equally large domains
above and below it, with the magnetization pointing in $\pm\hat x$
respectively.
In the inset to Fig.~\ref{fig_en} we compare the energy of the two-domain
state ($E_2$) with the single-domain (FM) state. Shown is the quantity
$E_{dw}=(E_2-E_{FM})/L_x$ which is a function of the
aspect ratio $r$ only. The formation of a domain wall defining the two-domain state is
energetically favored when $r<r_c\approx 4.8$.
For larger $r$, the formation of the domain wall is
energetically unfavorable. A distinguishing feature of the dipole lattice is that the energy of the domain
wall per unit length depends on the aspect ratio only, not on the
size of the sample. One does not see a proliferation of
domains with increasing array size.

We now examine the energy of a vortex state in a
rectangular sample. For the sake of simplicity,
we consider an elliptic vortex, as defined in note
\onlinecite{note_vortex}.
This configuration is not expected to be the `exact' 
ground state, because of corner effects.
A rigorous treatment should allow for spin relaxation,
which has been done in a few cases (see below).
However, it is not feasible to do that for a large
number of big samples of different aspect ratios.
We are confident that our main results, Eq.~(\ref{eq_en-vort})
and Figs. 1 (main) and 2, are robust, because corner
effects should be negligible in the limit of big sizes.

\begin{figure}
\includegraphics[width=7.5cm,angle=0,clip=true]{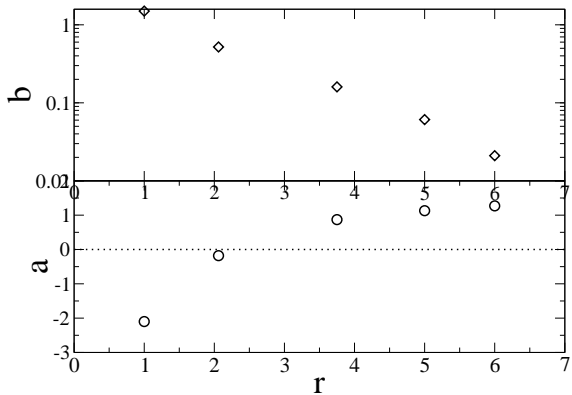}
\caption{Dependence of the parameters $a(r)$ (open circles) and $b(r)$ (open diamonds) 
on the aspect ratio $r$ of a rectangular sample. 
Their meaning is illustrated in Fig.~\protect\ref{fig_en}
(main).}
\label{fig_ab}
\end{figure}

In Fig.~\ref{fig_en} (main)
we show the energy $E_V$ of a vortex as a function of
the size $L_x$, for different aspect ratios $r$.
Unlike the domain wall energy, which depends only upon $r$, the  quantity $(E_V-E_{FM})/L_x$ depends
both on the shape and the size of the sample.
We find that the vortex energy obeys the functional form 
\be
(E_V-E_{FM})/L_x=a(r) -b(r)\ln L_x ,
\label{eq_en-vort}
\ee
with $a(r)$ and $b(r)$ plotted in Fig.~\ref{fig_ab}.
The first term, $a(r)$, does not depend on $L_x$ and displays an $r$-dependence 
which strongly resembles $E_{dw}(r)$ (see the inset
of Fig.~\ref{fig_en}). An average of the magnetization
of the vortex in the upper ($y>0$) and lower ($y<0$) regions 
approximates a two-domain
state, and $a(r)$ describes the interaction between
these upper and lower parts of the vortex.

The second term, $-b(r)\ln L_x$, is always negative and drives vortex formation.
Consider a two-domain ferromagnetic alignment as an initial condition
on the array. The energy gain for deformation into a vortex increases as $L_x$ is increased,
but strongly decreases as $r$ is increased.
Therefore,  provided that the size $L_x$ is sufficiently large,
 the vortex state appears to have an energy lower than parallel
aligned states for any aspect ratio. The size necessary for 
this depends on $r$.\cite{note_mv}

Our main study is
based on the direct numerical evaluations of dipolar energy (\ref{eq_energy})
for a number of fixed initial spin arrangements.
This method allows to determine the exact energies for
specified configurations in very large lattices.
In order to take into account the possibility of spin
relaxation, in a few cases
configurations that produce stable configurations have
also been found using numerical methods that adjust local spin
orientations in such a way as to minimize the total energy iteratively.
The results are generally consistent with the main study:
Vortex-like configurations appear for large enough 
arrays (large $L_x$), and can evolve into some type of 
domain structure that reduces magnetostatic energy near the
array edges.
For large values of the aspect ratio vortices do not form, but
domains form in proximity of the smallest sides because a FM state
would have the local magnetization perpendicular to them, which is
energetically unfavourable. These edge effects do not seem
to penetrate the whole sample for large enough arrays. 
The actual configuration is determined
by instabilities of parallel rectangular domains with respect to the
formation of flux-closure domains with triangular shape.
More general and precise conclusions would require a
systematic simulation work\cite{Dietrich} 
which is beyond the scope of this Rapid Note.

We now turn to approximations involving the range $R$ of the dipolar interaction. 
Truncation of dipole sums\cite{Fazekas_2003} 
by imposing finite range cutoffs for dipolar interactions is often 
used in numerical simulations in order to reduce computation times.
This may be a severe approximation for dipole lattices regardless of lattice size.

The stability of a ground state configuration can be examined by studying the frequencies of
spin wave excitations. 
It is widely accepted that, for an infinite range interaction,
the ferromagnetic state on a
triangular lattice is locally stable. In consequence, the linear spin wave frequency
spectrum is real over the entire Brillouin zone,
and $\omega(q)\approx \sqrt{q}$ at small $q$.
The non-analyticity of $\omega(q)$ in $q=0$ is due to the
long-range character of the interaction.

An estimate of the general behavior of spin wave frequencies in parallel (FM) spin
configuration under the finite range approximation can be made as follows.
If the spin configuration is assumed to consist of infinite parallel lines
it is easy to calculate the domain wall energy
because interaction between lines decays exponentially, and we get
$E_{dw}^{\infty}=0.2356$. This positive value means
that creating a domain wall in an infinite system with infinite range
interaction has a finite energy cost. Now let the range $R$ be finite.

The spin wave frequency has the following general form:\cite{Maleev}
$\omega(\vec q) =\sqrt{(A-B)(A+B)}$, where
\bea
A-B &=& \sum_i {1\over r_i^3}[ 2+\cos(\vec q\cdot\vec r_i)] \\
A+B &=& D_{yy}(\vec q) - D_{yy}(0)
\eea
with $D_{yy}(\vec q)=\sum_i {1\over r_i^3}\left(1-3\frac{y_i^2}{r_i^2}\right)
\cos(\vec q\cdot\vec r_i)$.

\noindent The quantity $(A-B)$ is manifestly and strictly positive, while $(A+B)$
may take both signs, as we are going to argue. 
For infinite range interactions,\cite{Maleev}
$D_{yy}(\vec q) - D_{yy}(0)\approx cq$, with a positive
prefactor $c$. For a finite range $R$ and small $q$, we can expand
the cosine in $D_{yy}$ and replace the discrete summation
on the angle $\theta$ [$\vec r_i\equiv (r_i,\theta_i)$]
with an integral, getting
\be
A+B \approx
\frac{\pi}{8}\sum_r\frac{q^2}{r}(6\sin^2\phi -1)
\label{apb}
\ee
where $\phi$ is the angle between $\vec q$ and the $x$ axis.

The above summation is limited by the range $R$ of the
interaction and it is valid for values of $q$ small enough, $qR <1$. 
Eq.~(\ref{apb}) shows that $A+B$ is negative,
therefore pointing out an instability, in the range
$\sin^2\phi <1/6$: $|\phi|<\phi_c=24.1^\circ$ and
$|\pi-\phi|<\phi_c~$ (see note \onlinecite{note_ins}).
It is worth noting that, with increasing $R$, $\phi_c$ keeps constant 
but the maximal value $q_M\approx 1/R$ at which $(A+B)<0$ decreases.
The conclusion is that, for any finite value of the
range $R$ of dipolar interaction, there is a portion
of the Brillouin zone where $\omega^2(\vec q) <0$.
This means that the ferromagnetic state is unstable for any finite $R$.

We have found numerically that the domain wall energy $E_{dw}(R)$, in the
case of finite interaction range $R$, has {\it negative} values for any $R$.
A limiting value exists
 of $E_{dw}(R\to\infty)=-0.3570$. This result means that domain wall
formation is energetically favored for any finite $R$, consistently with
the spin-wave instability discussed before.

Note that there is an apparent inconsistency here. 
Above we stated that $E_{dw}^\infty >0$. How is it possible that
$\lim_{R\to\infty} E_{dw}(R)=E_{dw}(\infty)<0$
and $E_{dw}^\infty >0$?
The answer follows from our previous analysis:
the domain wall energies $E_{dw}(\infty)$ and $E_{dw}^\infty$
differ because they correspond to different ways to take
the thermodynamic limit. $E_{dw}(\infty)$ is evaluated
assuming that each spin interacts with all spins within a circle
of increasing radius $R$, which corresponds to a system of
aspect ratio of order one. For this system (see the inset to Fig.~\ref{fig_en}) 
we find correctly a negative domain wall energy.
On the other hand, evaluating $E_{dw}^\infty$ with
the assumption that each spin interacts with an infinite number of other lines
(i.e. a system with very large aspect ratio),
we find a positive domain wall energy.
A quantitative understanding of the disagreement between
$E_{dw}(\infty)$ and $E_{dw}^\infty$ is also possible.\cite{note_edw}

The tendency to form domain walls and vortices is very sensitive to the presence of local
magnetic anisotropies. Our final considerations are for the effects of quadratic and quartic local anisotropies
on spin configurations in a dipole lattice. The anisotropies are defined by
\be
E^{ani} = -K_2\sum_i\cos^2\phi_i - K_4\sum_i (\sin^4\phi_i+\cos^4\phi_i)
\ee
\noindent where $\phi_i$ is the angle formed by spin $\vec S_i$ with the $x$ axis and
$K_2,K_4>0$ so that both single-site anisotropies and shape anisotropy
favor the $x$ axis. We have evaluated\cite{note_K} $E^{ani}$ for a vortex 
configuration, with the result that
\be
{1\over L_xL_y} (E^{ani}_V - E^{ani}_{FM}) =
K_2 {r\over 1+r} + K_4{r\over (1+r)^2}
\ee
\noindent which should be added to the vortex dipolar energy (per spin),
${1\over L_xL_y} (E^{dip}_V - E^{dip}_{FM}) =
{\Omega\over L_y} [ a(r) -b(r)\ln L_x]$.
For $r\simeq 1$, the vortex state is favored for $L_x < \bar c(\Omega/K)$,
with $K=\max\{K_2,K_4\}$ and $\bar c\approx 4/10$.
For large $r$, a very small anisotropy is enough to favor the
ferromagnetic state for any $L_x$.

As for future work, it would be interesting to perform more
sophisticated analyses of the ground state (e.g., simulated
annealing\cite{Dietrich}), allowing for spin relaxation and
overcoming of energy barriers. The comprehension of temperature
effects would be worth studying as well.


\begin{thebibliography}{99}
\bibitem{note_first-result}
This can be easily seen by rewriting $\vec H(P)$ in the form
$
H_{x,y}(P) = - \oint \nabla
\left( {\vec r\cdot\vec\epsilon_{x,y}\over r^3}\right)
\cdot\vec{d\ell} \equiv 0
$
where $\vec d\ell=d\ell\vec S$ is tangent to $\Gamma$ and
$\vec\epsilon_{x,y}$ are unit vectors.
\bibitem{note_second-result}
The proof is simple for a circle.
For any spin tangent to the curve at the angle
$\phi_0+\phi$ there is another spin tangent in $\phi_0-\phi$.
Evaluation of the dipolar field in $\phi_0$ shows immediately that it also is tangent
to the circle.
\bibitem{note_vortex}
Consider an ellipse concentric with the rectangle, with axes $r_x$ 
and $r_y$ parallel to the
cartesian axes and $\frac{r_x}{r_y}=r$. Now allow the size of the ellipse
to vary while holding $r$ constant.
Each spin is intersected for some value of $r_x$:
a vortex state is defined by tangential alignment of the 
spin at its intersection with the ellipse.
\bibitem{note_ins}
It can be shown numerically that this result is exact,
because lattice sites at any distance $R$ are
located on special symmetry points so that summing
on them is equivalent to integrate on the full circle.
\bibitem{note_edw}
The quantity
$\Delta E_{dw}^\infty = E_{dw}^\infty - E_{dw}(\infty)$
can be evaluated analytically. It amounts to
calculate the `complement' of $E_{dw}(R)$, i.e.
to sum the dipolar coupling betweens sites $i,j$ such that $r_{ij}>R$
and let $R$ going to infinity. 
We find $\Delta E_{dw}^\infty={16\over 27}$, which matches
very well the numerical result
$\Delta E_{dw}^\infty=0.2356+0.357=0.5926$.
\bibitem{note_K}
We have calculated analytically the averages
$\langle\cos^2\phi\rangle,\langle\cos^4\phi\rangle,\dots$
for a continuous field $\vec S(\vec r)$ along an ellipse $\Gamma$,
with $\vec S$ tangent to $\Gamma$.
\end{thebibliography}
\end{document}